\documentclass[9pt, sigconf]{IEEEtran}

\def\BibTeX{{\rm B\kern-.05em{\sc i\kern-.025em b}\kern-.08emT\kern-.1667em\lower.7ex\hbox{E}\kern-.125emX}}

\usepackage{url}
\usepackage{multirow}
\usepackage{subfigure}
\usepackage{epsfig}
\usepackage{graphicx}
\usepackage{amsmath}
\usepackage{amssymb}
\usepackage{color}
\usepackage{setspace}
\usepackage{cite}
\usepackage{hhline}   
\usepackage{algorithm}
\usepackage{algorithmic}
\usepackage{flushend}
\usepackage{graphics}
\usepackage{tikz}
\usepackage{array}
\usepackage{animate}

\definecolor{darkgreen}{RGB}{11,140,21}

\renewcommand{\baselinestretch}{0.94}
\begin{document}

\title{Thermal and IR Drop Analysis Using\\ Convolutional Encoder-Decoder Networks}

\author{
Vidya A. Chhabria$^1$, Vipul Ahuja$^2$, Ashwath Prabhu$^2$,  Nikhil Patil$^2$, Palkesh Jain$^2$, and Sachin S. Sapatnekar$^1$\\
\normalsize{$^1$University of Minnesota, USA;  $^2$Qualcomm Technologies Inc, India}
\vspace{-0.6cm}
}

\maketitle

\begin{abstract}  
Computationally expensive temperature and power grid analyses are required during the design cycle to guide IC design.
This paper employs encoder-decoder based generative (EDGe) networks to map these analyses to fast and accurate image-to-image and sequence-to-sequence translation tasks. The network takes a power map as input and outputs the corresponding temperature or IR drop map. We propose two networks: (i) ThermEDGe: a static and dynamic full-chip temperature estimator and (ii) IREDGe: a full-chip static IR drop predictor based on input power, power grid distribution, and power pad distribution patterns. The models are design-independent and must be trained just once for a particular technology and packaging solution. ThermEDGe and IREDGe are demonstrated to rapidly predict on-chip temperature and IR drop contours in milliseconds (in contrast with commercial tools that require several hours or more) and provide an average error of 0.6\% and 0.008\% respectively.
\end{abstract}

\bstctlcite{IEEEexample:BSTcontrol}
\section{Introduction}
\label{sec:intro}

\noindent
One of the major challenges faced by an advanced-technology node IC designer is
the overhead of large run-times of analysis tools. Fast and accurate analysis tools that aid quick design
turn-around are particularly important for two critical, time-consuming simulations that are performed
several times during the design cycle:
\begin{itemize}
\item {\em Thermal analysis}, which checks the feasibility of a
placement/floorplan solution by computing on-chip temperature distributions in
order to check for temperature hot spots.
\item {\em IR drop analysis} in power distribution networks (PDNs), which
diagnoses the goodness of the PDN by determining voltage (IR) drops from the
power pads to the gates.
\end{itemize}   
The underlying computational engines that form the crux of both analyses are
similar: both simulate networks of conductances and current/voltage
sources by solving a large system of equations of the form $G {\bf V} = {\bf J}$~\cite{Zhan08,Zhong05} with millions to billions of
variables.  In modern
industry designs, a single full-chip temperature or IR drop simulation can take hours to several
hours.
Accelerating these analyses opens the door to optimizations in the design cycle that
iteratively invoke these engines under the hood. 

The advent of machine learning (ML) has presented fast and fairly accurate solutions to these
problems~\cite{zhang18, juan12,
tan19, Lin18,incpird, powernet} which can successfully be used in early design cycle optimizations, operating within larger allowable error margins at these stages.
To the best of our knowledge, no published
work addresses full-chip ML-based thermal analysis: the existing literature
focuses on coarser-level thermal modeling at the system level~\cite{zhang18,
juan12, tan19}.  For PDN analysis, the works in~\cite{Lin18, incpird} address
incremental analysis, and are not intended for full-chip
estimation.  The work in~\cite{powernet} proposes
a convolutional neural network (CNN)-based implementation for full-chip IR drop
prediction, using cell-level power maps as features. However, it assumes
similar resistance from each cell to the power pads, which may not be valid for practical power
grids with irregular grid density.  The analysis divides the chip into regions ({\it
tiles}), and the CNN operates on each tile and its near
neighbors.  Selecting an appropriate tile and window size is nontrivial -- small windows could violate the principle of
locality~\cite{Chiprout04}, causing inaccuracies, while large windows could
result in large models with significant runtimes for training and inference.
Our approach bypasses 
window size selection by providing the entire power map
as a feature, allowing ML to learn the window size for accurate estimation.


We translate static analysis problems to an image-to-image translation task and
dynamic analysis problems to video-to-video translation, where the inputs are
the power/current distributions and the required outputs are the temperature or
IR drop contours. For static analysis, we employ fully convolutional (FC) EDGe
networks for rapid and accurate thermal and IR drop analysis. FC EDGe networks
have proven to be very successful with image-related problems with 2-D
spatially distributed data~\cite{fcn, unet, mao16, segnet} when compared to
other networks that operate without spatial correlation
awareness.  For transient analysis, we use long-short-term-memory (LSTM)
based EDGe networks that maintain memory of analyses at prior time steps.

\begin{figure}[tb]
\centering
\includegraphics[width=0.43\textwidth]{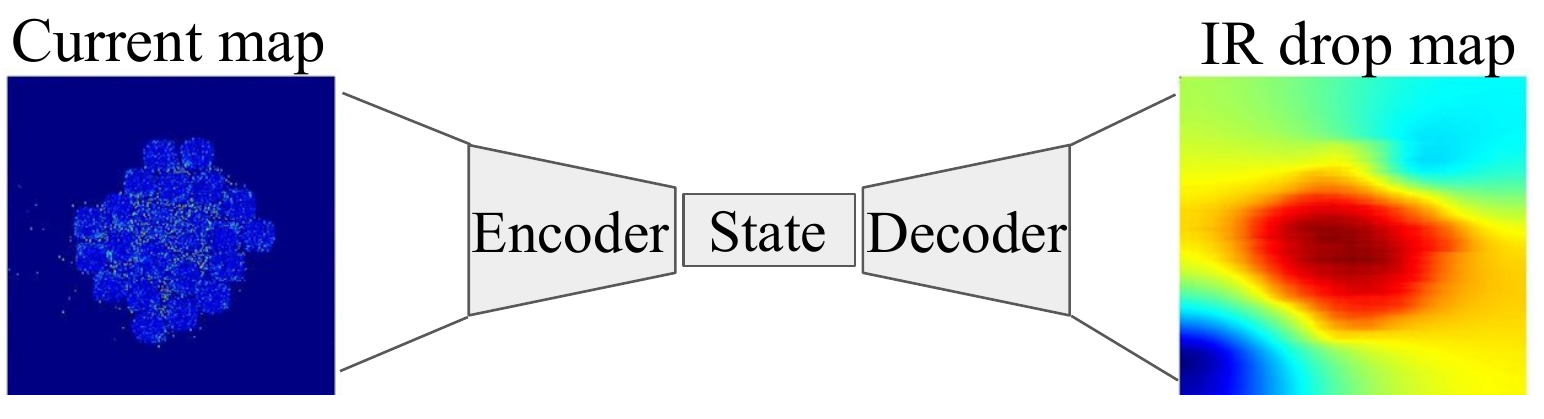}
\caption{Image-to-image translation using EDGe network.}
\label{fig:enc-dec-arch}
\vspace{-1.5em}
\end{figure}

Based on these concepts, this work proposes two novel ML-based analyzers: {\bf
ThermEDGe} for both \textit{full-chip} static and transient thermal analysis, and {\bf IREDGe} for
\textit{full-chip} static IR drop estimation. The fast inference times of ThermEDGe and IREDGe enable {\bf full-chip} thermal and IR drop analysis in milliseconds, as opposed to runtimes of several hours using
commercial tools. We obtain average error of 
0.6\% and 0.008\% for ThermEDGe and IREDGe, respectively, over a range of testcases. {\em We will open-source our software.}

Fig.~\ref{fig:enc-dec-arch} shows a general top-level structure of an EDGe
network. It consists of two parts: (i) the encoder/downsampling path, which
captures global features of the 2-D distributions of power dissipation, and
produces a low-dimensional state space and (ii) the decoder/upsampling path,
which transforms the state space into the required detailed outputs
(temperature or IR drop contours). 
The EDGe network is well-suited for PDN/thermal
analyses because: \\
{\em (a)} The convolutional nature of the encoder 
{\em captures the dependence of both problems on the spatial distributions of power}.
Unlike CNNs, EDGe networks contain a decoder which acts as
a generator to convert the extracted power and PDN density features into
accurate high-dimensional temperature and IR drop contours across the chip. \\
{\em (b)} The trained EDGe network model for static analysis is
{\em chip-area-independent}: it only stores the weights of the convolutional kernel,
and the same filter can be applied to a chip of any size.  The selection of the
network topology (convolution filter size, number of convolution layers) is
related to the expected sizes of the hotspots rather than the size of the chip:
these sizes are generally similar for a given application domain, technology,
and packaging choice. \\
{\em (c)} Unlike prior methods~\cite{powernet} that operate tile-by-tile, where finding the right tile and window size for accurate analysis
is challenging, {\em the choice of window size is treated as an ML
hyperparameter tuning problem} to decide the necessary amount of input spatial information.

\section{EDGe Network for PDN and Thermal Analysis}

\subsection{Problem formulations and data representation}
\label{sec:problem-def}

\begin{figure}[tb]
\centering
\includegraphics[width=8.8cm]{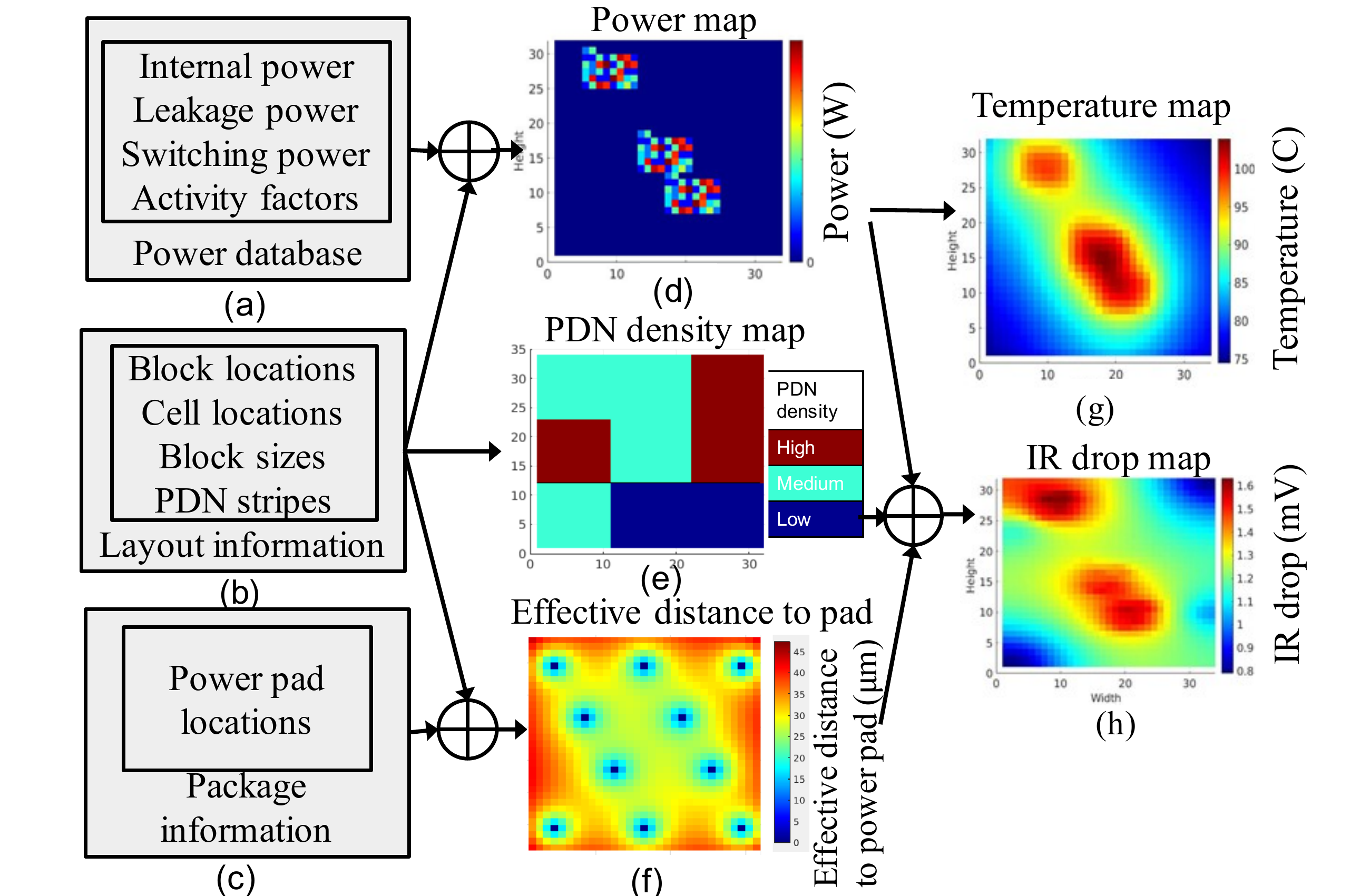}
\caption{Data representation: Mapping PDN and thermal analysis problems into
image-to-image translations tasks.}
\label{fig:data-rep}
\vspace{-1em}
\end{figure}

\noindent
This section presents the ML-based framework for ThermEDGe and IREDGe.  The
first step is to extract an appropriate set of features from a standard
design-flow environment.  The layout database provides the
locations of each instance and block in the layout, as
outlined in Fig.~\ref{fig:data-rep}(a).  This may be combined with information
from a power analysis tool such as~\cite{voltus} (Fig.~\ref{fig:data-rep}(b))
that is used to build a 2-D spatial power map over the die area.  

For thermal analysis using ThermEDGe, both the inputs and outputs are images
for the static case, and a sequence of images for the transient case.  Each
input image shows a 2-D die power distribution (static) image, and each output
image is a temperature map across the die (Fig.~\ref{fig:data-rep}). For static PDN analysis, the output is an IR drop map across the full chip. However, in addition to the 2-D power distributions, IREDGe has two other inputs:

\noindent
(i)~\underline{\it A PDN density map}: This feature is generated by extracting the average PDN 
pitch in each region of the chip. For example, when used in conjunction with the PDN styles in~\cite{jsingh,OpeNPDN}, where the chip uses regionwise uniform PDNs, the
average PDN density in each region, across all metal layers, is provided as an input (Fig.~\ref{fig:data-rep}(e)).

\noindent
(ii)~\underline{\it An effective distance to power pad}: This feature represents the equivalent distance from an instance to all power pads in the package. We compute the effective distance of each instance, $d_e$, to $N$ power pads on the chip as the harmonic sum of the distances to the pads:
\begin{equation}
    d_e^{-1} = d_1^{-1} + d_2^{-1} + ...+d_N^{-1}
\label{eq:deffective}
\end{equation}
where $d_i$ is the distance of the $i^{th}$ power pad from the instance. Intuitively, the effective distance metric and the PDN density map together, represent the equivalent resistance between the instance and the pad. The equivalent resistance is a parallel combination of each path from the instance to the pad. We use distance to each pad as a proxy for the resistance in Eq.~\eqref{eq:deffective}. Fig.~\ref{fig:data-rep}(f) shows a typical ``checkerboard'' power pad layout for flip-chip packages~\cite{checkerboard1, checkerboard2}.  

Temperature depends on
the ability of the package and system to conduct heat to the ambient, and IR drop depends on off-chip (e.g., package) parasitics. In this work, our focus is strictly on-chip, and 
both ThermEDGe and IR-EDGe are trained for fixed models of a given technology, package, and system.

Next, we map these
problems to standard ML networks:  
\begin{itemize}
\item
For static analysis, the problem formulations require a translation from an
input power image to an output image, both corresponding to contour maps over the
same die area, and we employ a {\em U-Net-based EDGe network}~\cite{unet}.
\item
The dynamic analysis problem requires the conversion of a sequence of input
power images, to a sequence of output images of temperature contours, and this problem is
addressed using an {\em LSTM-based EDGe network}~\cite{seq2seq}.
\end{itemize}
We describe these networks in the rest of this section.

\begin{figure}[htb]
\centering
\includegraphics[width=8.8cm]{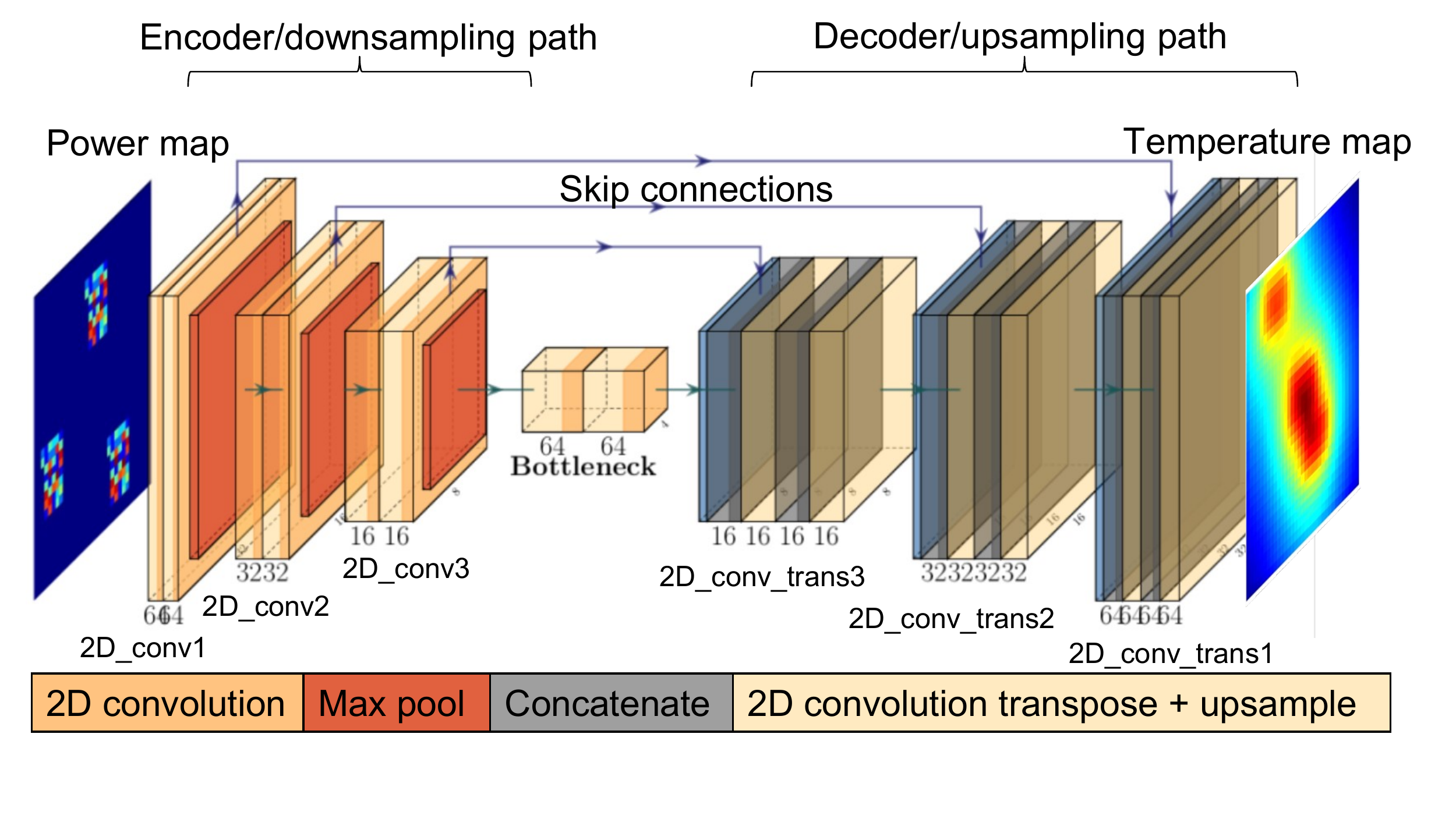}
\caption{U-Net-based EDGe network for static thermal and PDN analysis.}
\label{fig:thermedge}
\vspace{-1.5em}
\end{figure}

\subsection{U-Nets for static thermal and PDN analysis}
\label{sec:unet}

\subsubsection{Overview of U-Nets}

CNNs are successful in extracting 2-D spatial information for image
classification and image labeling tasks, which have low-dimensional outputs (class or label). For PDN and thermal analysis tasks, the required outputs
are high-dimensional distributions of IR drop and temperature
contour, where the dimensionality corresponds to the number of
pixels of the image and the number of pixels 
is proportional to the size of the chip. This calls for a generator network
that can translate the extracted low-dimensional power and PDN features from a
CNN-like encoder back into high-dimensional representing the required output data. 

Fig.~\ref{fig:thermedge} shows the
structure of the EDGe network used for static PDN and thermal analysis.
At the top level, it consists of two networks:

\noindent
(a)~\underline{\em Encoder/downsampling network} Like a CNN, the network utilizes a sequence of 2-D
convolution and max pooling layer pairs that extract key features from the
high-dimensional input feature set.  The convolution operation performs a
weighted sum on a sliding window across the image~\cite{conv-deconv}, and the
max pooling layer reduces the dimension of the input data by extracting the
maximum value from a sliding window across the input image. In
Fig.~\ref{fig:thermedge}, the feature dimension is halved at each stage by each
layer pair, and after several such operations, an encoded, low-dimensional,
compressed representation of the input data is obtained. For this reason, the
encoder is also called the downsampling path: intuitively, downsampling
helps understand the {\em ``what"} (e.g., ``Does the image contain power or IR
hotspots?'') in the input image but tends to be imprecise with the {\em ``where"}
information (e.g., the precise locations of the hotspots).  The latter is
recovered by the decoder stages.

\noindent
(b)~\underline{\em Decoder/upsampling network} Intuitively, the generative decoder is
responsible for retrieving the {\em ``where"} information that was lost during
downsampling, This distinguishes an EDGe network from its CNN counterpart.  The
decoder is implemented using the transpose convolution~\cite{conv-deconv} and
upsampling layers. Upsampling layers are functionally the opposite of a
pooling layer, and increase the dimension of the input data matrix by replicating the rows and columns.  

\subsubsection{Use of skip connections}

Static IR drop and temperature are strongly correlated to the input power -- 
a region with high power on the chip could potentially have an IR or temperature
hotspot in its vicinity.  U-Nets~\cite{unet} utilize {\it skip} connections
between the downsampling and upsampling paths, as shown in
Fig.~\ref{fig:thermedge}.  These connections take information from one layer
and incorporate it using a {\it concatenation} layer at a deeper stage skipping intermediate layers, and appends it to the embedding
along the z-dimension.  

For IR analysis, skip connections combine the local power, PDN information, and power pad locations from the
downsampling path with the global power information from the upsampling path,
allowing the underlying input features to  
and directly shuttle to the layers closer to the output, and are similarly helpful for thermal analysis.  This helps
recover the fine-grained ({\em ``where"}) details that are lost in the encoding part
of the network (as stated before) during upsampling in the decoder for detailed temperature and IR
drop contours.

\subsubsection{Receptive fields in the encoder and decoder networks}

The characteristic of PDN and thermal analyses problems is that the IR drop and
temperature at each location depend on both the local and global power
information. 
During convolution, by sliding averaging windows of an appropriate size
across the input power image, the network captures local spatially correlated
distributions. For capturing the larger global impact of power on temperature
and IR drop, max pooling layers are used after each convolution to
appropriately increase the size of the {\it receptive field} at each stage of
the network. The {\it receptive field} is defined as the region in the input
2-D space that affects a particular pixel, and it determines the impact
of  the local, neighboring, and global features on PDN and thermal
analysis.  

In a deep network, the value of each pixel feature is
affected by all of the other pixels in the receptive field at the previous convolution stage, with the largest contributions coming from pixels near the center
of the receptive field.  Thus, each feature not only captures its receptive field in the input image, but also gives an exponentially higher weight to the middle of that region~\cite{receptive-field}. This matches
with our applications, where both thermal and IR maps for a pixel are most
affected by the features in the same pixel, and partially by features in
nearby pixels, with decreasing importance for those that are farther away.
The size of the receptive field at each stage in the network is determined by the convolutional filter size, number of convolutional
layers, max pooling filter sizes, and number of max pooling layers.

On both the encoder and decoder sides in Fig.~\ref{fig:thermedge}, we use three
stacked convolution layers, each followed by 2$\times$2 max-pooling to
extract the features from the power and PDN density images. The number of layers and filter sizes are determined based on the magnitude of the hotspot 
size encountered during design iterations.

\subsection{LSTM-based EDGe network for transient thermal analysis}

Long short term memory (LSTM) based EDGe networks are a special kind of recurrent neural
network (RNN) that are known to be capable of learning long term dependencies
in data sequences, i.e., they have a memory component and are capable of learning
from past information in the sequence. 

\begin{figure}[h]
\centering
\includegraphics[width=0.47\textwidth]{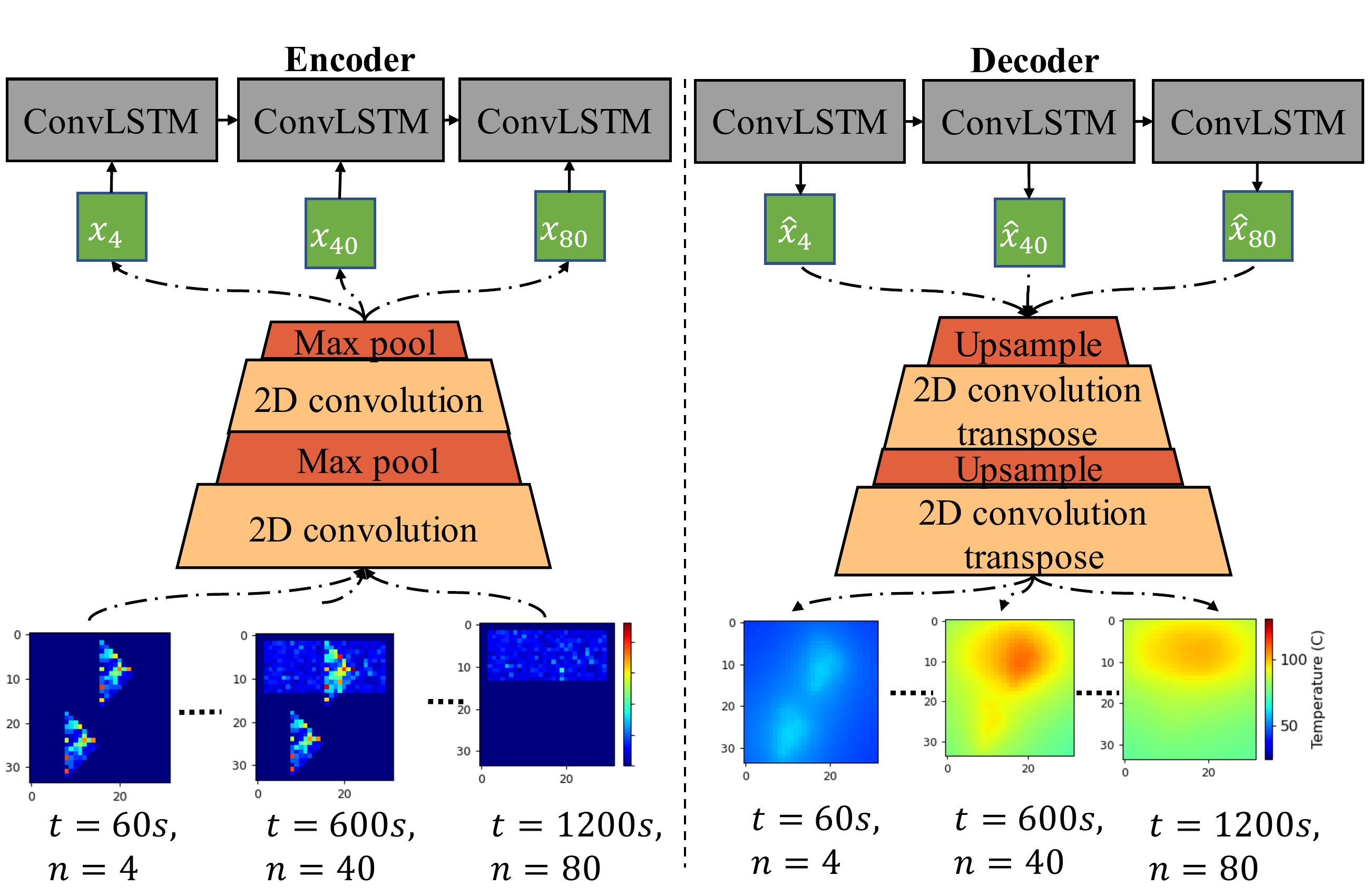}
\caption{LSTM-based EDGe network for transient analysis in ThermEDGe.}
\label{fig:lstm-edge}
\end{figure}

\begin{figure}[h]
\centering
\includegraphics[width=8.5cm]{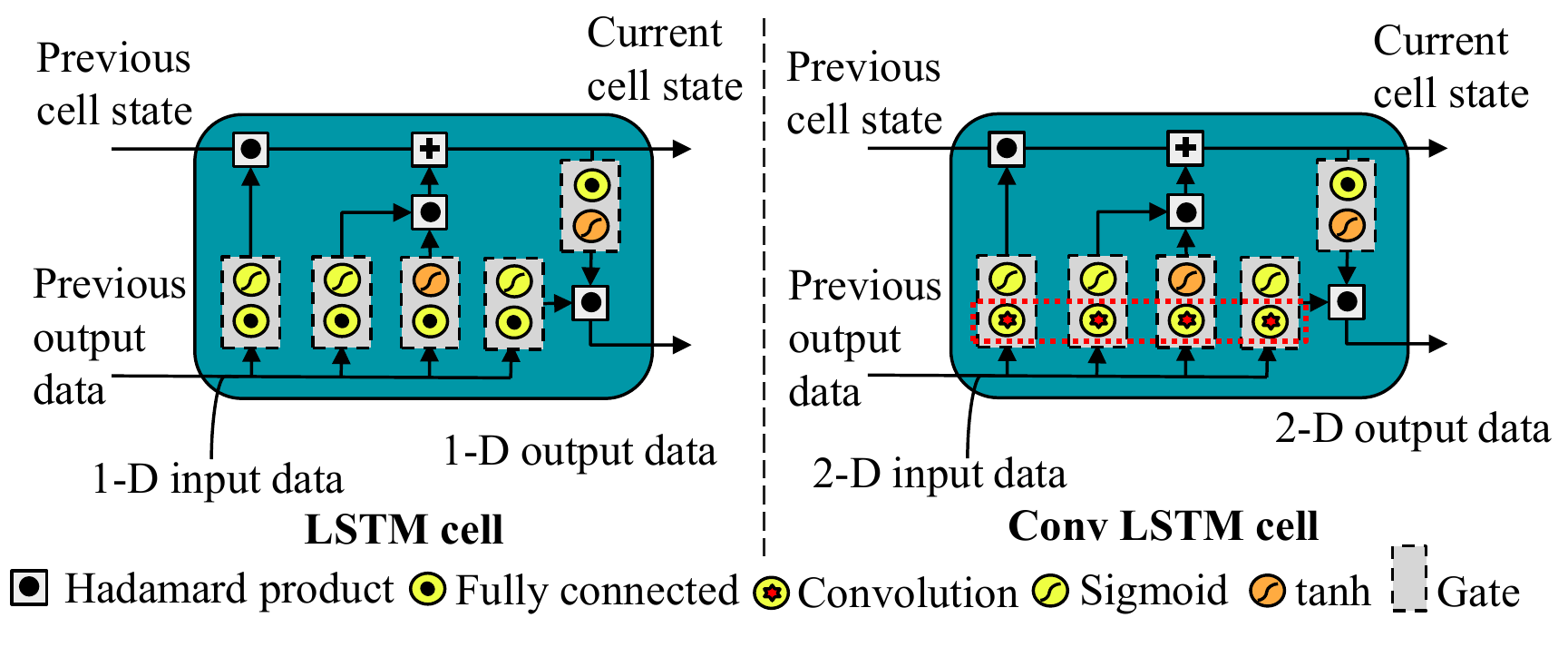}
\caption{A fully connected LSTM cell (left) and a ConvLSTM cell (right).}
\label{fig:conv-lstm}
\end{figure}

For transient thermal analysis,  the structure of ThermEDGe is shown in Fig.~\ref{fig:lstm-edge}. The core architecture 
is an EDGe network, similar to the static analysis problem described in Section~\ref{sec:unet}, except that  the 
network uses additional LSTM cells to account for the time-varying component. The figure demonstrates the time-unrolled LSTM where input power
frames are passed to the network one frame at a time. The LSTM cell accounts
for the history of the power maps to generate the output temperature frames for all time steps.  The network is used for  sequence-to-sequence translation in transient thermal analysis, where the input is a set of time-varying power maps and the output is a set of time-varying temperature maps (Section~\ref{sec:problem-def}).

Similar to the static ThermEDGe network (Fig.~\ref{fig:thermedge}), the encoder consists of convolution and max pooling layers to downsample and extract critical local and global spatial information and the decoder consists of upsampling and transpose convolution layers to upsample the encoded output. 
However, in addition, transient ThermEDGe has LSTM layers in both the encoding and decoding paths.

A standard LSTM cell is shown in Fig.~\ref{fig:conv-lstm} (left). 
While the basic LSTM cell uses fully connected layers within each gate, our application
uses a variation of an LSTM cell called a convolutional LSTM
(ConvLSTM)~\cite{conv-lstm}, shown in Fig.~\ref{fig:conv-lstm} (right).  In
this cell, the fully connected layers in each gate are replaced by convolution
layers that capture spatial information.  Thus, the LSTM-based EDGe network
obtains a spatiotemporal view that enables accurate inference.

\section{ThermEDGe and IREDGe Model Training}

\noindent
We train the models that go into ThermEDGe and IREDGe to learn the temperature
and IR contours from the ``golden" commercial tool-generated or ground truth
data.  We train ThermEdge using the full physics-based thermal simulations from
the Ansys-Icepak~\cite{icepak} simulator, incorporating off-chip thermal
dynamics from package and system thermal characteristics.  IREDGe is trained
using static IR drop distribution from a PDN analyzer~\cite{pdnsim, voltus} for
various power, PDN density, and power pad distributions.

\subsection{Generating training data}
\label{sec:train}

\noindent
{\bf Static ThermEDGe and IREDGe}
A challenge we faced to evaluate our experiments is the dearth of public domain benchmarks that fit these applications. The IBM benchmarks~\cite{ibm}, are potential candidates for our applications, but they assume constant currents per region and represent an older technology node. Therefore, we generate our 
dataset which comprises of 50 industry-relevant testcases, where each testcase represents industry-standard workloads for commercial designs implemented in a FinFET technology. The power images of size
34$\times$32 pixels, with each pixel representing the power/temperature a
250$\mu$m$\times$250$\mu$m tile on an 8.5mm$\times$8mm chip.\footnote{Note that although the temperature and power map work at this resolution, the actual simulation consists of millions of nodes; using fewer node (e.g., one node per pixel) is grossly insufficient for accuracy.}  Our training is
specific to the resolution: for another image resolution, the model must be
retrained. We reiterate that although the training is performed on chips of fixed size, as
we show (Section~\ref{sec:results}), inference can be performed on a chip of any size as long as the
resolution remains the same.

For static ThermEDGe our training data is based on static Ansys-Icepak~\cite{icepak} 
simulations of these 50 testcases. For IREDGe, we synthesize irregular PDNs of
varying densities for each dataset element using {\it PDN templates}, as
defined by OpeNPDN~\cite{OpeNPDN}. These templates are a set of PDN building
blocks, spanning multiple metal layers in a 14nm commercial FinFET technology, which vary in their metal
utilization. For our testcases, we use three templates (high, medium, and low
density) and divide the chip into nine regions. As outlined in Section~\ref{sec:problem-def}), we use a checkerboard pattern of power pads that vary in the bump pitch and offsets across the dataset.

The synthesized full-chip PDN, power pad locations, and power distributions are taken as inputs into
the IR analyzer~\cite{pdnsim} to obtain  training data for IREDGe. For
each of the 50 testcases, we synthesize 10  patterns of PDN densities, and for each combination of combination of power and PDN distribution we synthesize 10 patterns of power pad distributions, creating a dataset with 5000 points.

\noindent
{\bf Transient ThermEDGe} 
For the transient analysis problem, our training data is based on transient
Ansys-Icepak~\cite{icepak} simulations. The size of the chip is the same as that of the static ThermEDGe testcases.
For each testcase, we generate 45 time-step simulations that range from 0 to
3000s, with irregular time intervals from the thermal simulator. Each
simulation is expensive in terms of the time and memory
resources: one simulation of  a 3000s time interval with 45
time-steps can take 4 hours with 2 million nodes. 
Transient ThermEDGe is trained using constant time steps of 15s which enables
easy integration with existing LSTM architectures which have an implicit
assumption of uniformly distributed time steps, without requiring additional
features to account for the time. 
The model is trained on 150 testcases with time-varying workloads as features, and their
time-varying temperature from Ansys-Icepak as labels.
\vspace{-1.0em}
\subsection{Model training}
\label{sec:training}
\noindent
For the static analysis problem, ThermEDGe and IREDGe use a 
static power map as input and PDN density map (for IR analysis only) to predict
the corresponding temperature and IR drop contours.  For the transient thermal
analysis problem, the input is a sequence of 200 power maps and the output is a
sequence of 200 temperature contours maps at a 15s time interval.  The ML model
and training hyperparameters used for these models are listed in
Table~\ref{tbl:parameters}.

\begin{table}[h]
\centering
\caption{ThermEDGe and IREDGe ML hyperparameters}
\label{tbl:parameters}
\resizebox{\linewidth}{!}{%
\begin{tabular}{||l||l|l||l|l|l||} 
\hhline{|t:===:t:===:t|}
\multicolumn{3}{||l||}{ML hyperparameters} & \begin{tabular}[c]{@{}l@{}}Static\\ThermEDGe \end{tabular} & IREDGe & \begin{tabular}[c]{@{}l@{}}Transient\\ThermEDGe\end{tabular} \\ 
\hhline{|:=:t:==::===:|}
\multirow{9}{*}{\begin{tabular}[c]{@{}l@{}}Model layer \\parameters \end{tabular}} & \multirow{2}{*}{\begin{tabular}[c]{@{}l@{}}2D\_conv1 \\2D\_conv\_trans1 \end{tabular}} & filter size & 5x5 & 3x3 & 5x5 \\ 
\cline{3-6}
 &  & \# filters & 64 & 64 & 64 \\ 
\cline{2-6}
 & \multirow{2}{*}{\begin{tabular}[c]{@{}l@{}}2D\_conv2 \\2D\_conv\_trans2 \end{tabular}} & filter size & 3x3 & 3x3 & 3x3 \\ 
\cline{3-6}
 &  & \# filters & 32 & 32 & 32 \\ 
\cline{2-6}
 & \multirow{2}{*}{\begin{tabular}[c]{@{}l@{}}2D\_conv3 \\2D\_conv\_trans3 \end{tabular}} & filter size & 3x3 & 3x3 & -- \\ 
\cline{3-6}
 &  & \# filters & 16 & 16 & -- \\ 
\cline{2-6}
 & Max pool layers & filter size & 2x2 & 2x2 & 2x2 \\ 
\cline{2-6}
 & \multirow{2}{*}{ConvLSTM} & filter size & -- & -- & 7x7 \\ 
\cline{3-6}
 &  & \# filters & -- & -- & 16 \\ 
\hhline{|:=::==::===:|}
\multirow{8}{*}{\begin{tabular}[c]{@{}l@{}}Training \\parameters \end{tabular}} & \multicolumn{2}{l||}{Epochs} & \multicolumn{3}{l||}{500} \\ 
\cline{2-6}
 & \multicolumn{2}{l||}{Optimizer} & \multicolumn{3}{l||}{ADAM} \\ 
\cline{2-6}
 & \multicolumn{2}{l||}{Loss function} & \multicolumn{3}{l||}{Pixelwise MSE} \\ 
\cline{2-6}
 & \multicolumn{2}{l||}{Decay rate} & \multicolumn{3}{l||}{0.98} \\ 
\cline{2-6}
 & \multicolumn{2}{l||}{Decap steps} & \multicolumn{3}{l||}{1000} \\ 
\cline{2-6}
 & \multicolumn{2}{l||}{Regularizer} & \multicolumn{3}{l||}{L2} \\ 
\cline{2-6}
 & \multicolumn{2}{l||}{Regularization rate} & \multicolumn{3}{l||}{1.00E-05} \\ 
\cline{2-6}
 & \multicolumn{2}{l||}{Learning rate} & \multicolumn{3}{l||}{1.00E-03} \\
\hhline{|b:=:b:==:b:===:b|}
\end{tabular}
}
\end{table}

We split the data in each set, using 80\% of the data points for
training, 10\% for test, and 10\% for validation.  The training dataset is
normalized  by subtracting the mean and dividing by the standard deviation. The
normalized golden dataset is used to train the network using an ADAM
optimizer~\cite{adam} where the loss function is a pixel-wise mean square error
(MSE).  The convolutional operation in the encoder and the transpose
convolution in the decoder are each followed by ReLU activation to add
non-linearity and L2 regularization to prevent over fitting.
The model is trained in Tensorflow 2.1 on an NVIDIA GeForce RTX2080Ti GPU.
Training run-times are: 30m each for static ThermEDGe and IREDGe, and
6.5h for transient ThermEDGe. 
We reiterate that this is
a one-time cost for a given technology node
and package, and this cost is amortized over
repeated use over many design iterations for multiple chips.


\section{Results and Analysis using TherEDGe/IREDGe}
\label{sec:results}
\subsection{Experimental setup and metric definitions}
\noindent
ThermEDGe and IREDGe are implemented using Python3.7 within a Tensorflow 2.1
framework.  We test the performance of our models on the 10\% of datapoints
reserved for the testset (Section~\ref{sec:training}) which are labeled T1--T21.  As mentioned earlier in Section~\ref{sec:train}, due the unavailability of new, public domain benchmarks to evaluate our experiments, 
we use benchmarks that 
represent commercial industry-standard design workloads. 

\noindent
{\bf Error metrics} As a measure of goodness of ThermEDGe and IREDGe predictions, we define a discretized regionwise error, $T_{err}~=~\left | T_{true} - T_{pred} \right | $, where $T_{true}$ is ground truth image, generated by commercial tools, and $T_{pred}$ the predicted image, generated by ThermEDGe. $IR_{err}$ is computed in a similar way. We report the average and maximum values of $T_{err}$ and $IR_{err}$ for each testcase. In addition, the percentage mean and maximum error are listed as a fraction of a temperature corner, i.e., 105$^\circ$C for thermal analysis and as a fraction of VDD$=0.7$V for IR drop analysis.

\subsection{Performance of ThermEDGe and IREDGe: Accuracy and speed}
\noindent
\noindent
{\bf Static ThermEDGe results} 
A comparison between the commercial tool-generated temperature and the
ThermEDGe-generated temperature map for T1--T5 are listed in
Table~\ref{tbl:thermal-results}.  The runtime of static ThermEDGe for
each the five testcases which are of size 34$\times$32 is approximately 1.1ms in
our environment. On average across the five testcases (five rows of the table), ThermEDGe has an average $T_{err}$ of 0.63$^\circ$C and a maximum $T_{err}$ of 2.93$^\circ$C.\footnote{Achieving this accuracy requires much finer discretization in Icepak.} These numbers are a small fraction when compared to the maximum ground truth temperature of these testcases (85 -- 150$^\circ$C).
The fast runtimes imply that our method can be used in the
inner loop of a thermal optimizer, e.g., to evaluate various chip configurations 
under the same packaging solution (typically chosen early in the
design process).  For such applications, this level of error is very
acceptable.

\begin{table}[h]
\centering
\caption{Summary of ThermEDGe results for static and transient analysis across 10 testcases.}
\label{tbl:thermal-results}
\resizebox{\linewidth}{!}{%
\begin{tabular}{||l|l|l||l|l|l||} 
\hhline{|t:===:t:===:t|}
\multicolumn{3}{||c||}{{\bf Static ThermEDGe
}} & \multicolumn{3}{|c||}{{\bf Transient ThermEDGe}} \\ 
\hhline{|:===::===:|}
\textbf{\#Testcase} &  \textbf{Avg. $\bf T_{err}$} & \textbf{Max $\bf T_{err}$} & \textbf{\#Testcase} &  \textbf{Avg. $\bf T_{err}$} & \textbf{Max $\bf T_{err}$}  \\
\hline \hline
T1 &  0.64C (0.61\%) & 2.76C (2.63\%) & T6 & 0.51C (0.49\%) & 5.59C (5.32\%)  \\ 
\hline
T2 &  0.63C (0.60\%) & 2.67C (2.54\%) & T7 & 0.58C (0.55\%) & 6.17C (5.88\%)  \\ 
\hline
T3 &  0.65C (0.62\%) & 2.93C (2.79\%) & T8 & 0.57C (0.54\%) & 5.83C (5.55\%)  \\ 
\hline
T4 &  0.48C (0.46\%) & 2.22C (2.11\%) & T9 & 0.52C (0.50\%) & 6.32C (6.02\%)  \\ 
\hline
T5 &  0.75C (0.71\%) & 2.86C (2.72\%) & T10 & 0.56C (0.53\%) & 7.14C (6.80\%) \\
\hhline{|b:===:b:===:b|}
\end{tabular}
}
\end{table}

\begin{figure}[h]
\centering
\includegraphics[width=6.5cm]{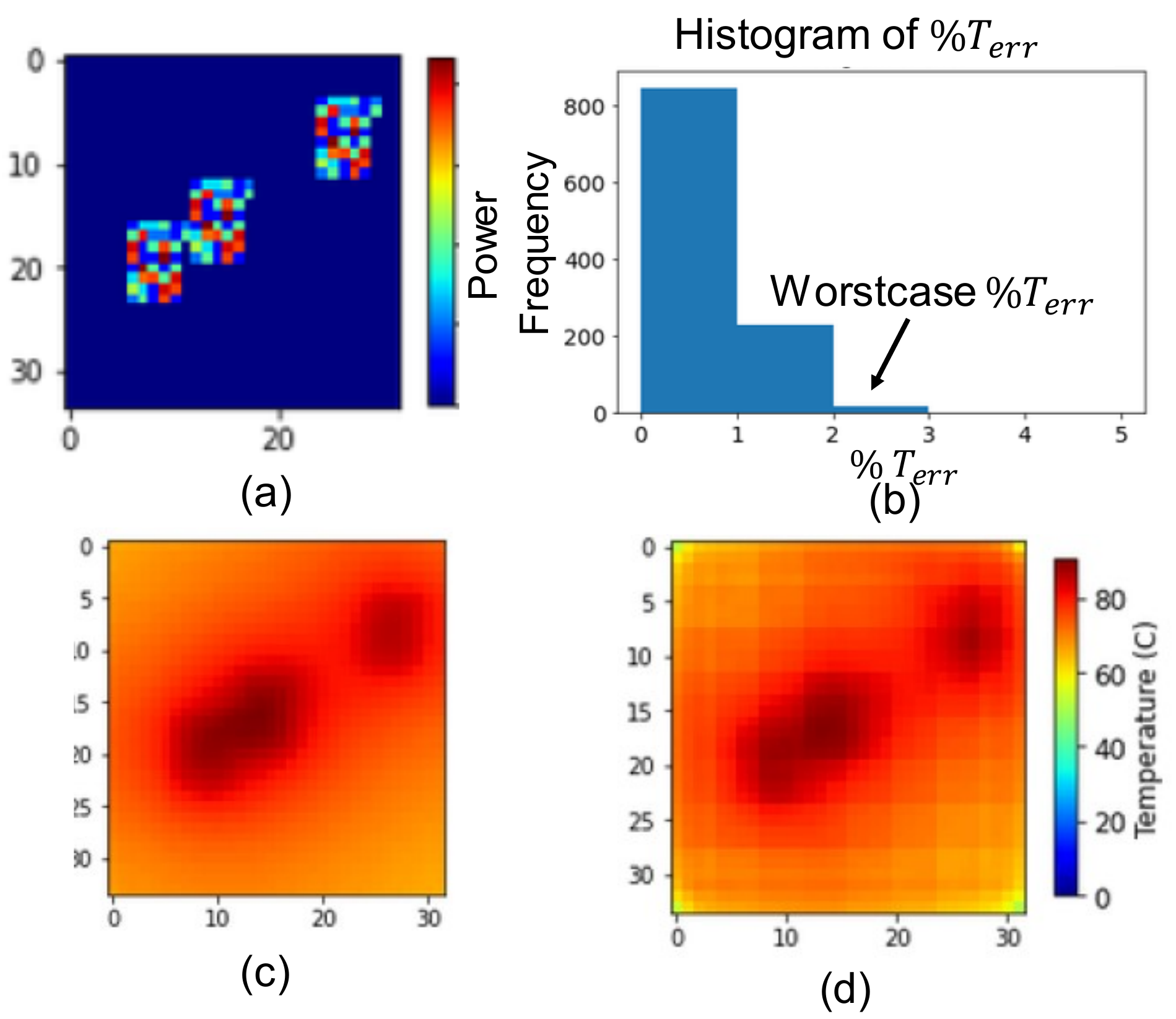}
\caption{ThermEDGe static temperature estimation on T1: (a) input normalized power distribution, (b) histogram of $T_{err}$ where maximum error is 2.76$^\circ$C which is very small compared to the maximum temperature of 85$^\circ$C, (c) ground truth temperature map, and (d) predicted temperature map.}
\label{fig:static-thermal-results}
\end{figure}

A graphical view of the predicted map for T1 is
depicted in Fig.~\ref{fig:static-thermal-results}. For a given input power
distribution in Fig.~\ref{fig:static-thermal-results}(a),  ThermEDGe generates
the temperature contour plots, as shown in
Fig.~\ref{fig:static-thermal-results}(d). We compare the predicted value against the true value
(Fig.~\ref{fig:static-thermal-results}(c)). The discrepancy is visually seen to
be small. Numerically, the histogram in Fig.~\ref{fig:static-thermal-results}(b) shows the distribution of \%$T_{err}$ across regions (Fig.~\ref{fig:static-thermal-results}(b). The average $T_{err}$ 0.64$^\circ$C and the maximum $T_{err}$ is 2.93$^\circ$C. This corresponds an average error of 0.52\% and worst-case error of 2.79\% as shown in the figure.

\noindent
{\bf Transient ThermEDGe results}
The transient thermal analysis problem is a sequence-to-sequence prediction
task  where each datapoint in the testset
has 200 frames of power maps at a 15s interval. Trained transient ThermEDGe
predicts the output temperature sequence for the input power sequence. 
We summarize the results in Table~\ref{tbl:thermal-results}.  The
inference run-times of T6--10 to generate a sequence 200 frames of
temperature contours is approximately 10ms in our setup.
Across the five testcases, the prediction has an average $T_{err}$ of 0.52\%
and a maximum $T_{err}$ of 6.80\% as shown. The maximum $T_{err}$ in our testcases occur during transients which do not have long-last effects (e.g., on IC reliability).  These errors are reduced to the average $T_{err}$ values at sustained peak temperatures.

\begin{figure}[ht]
    \centering
    \includegraphics[width=1.05\linewidth]{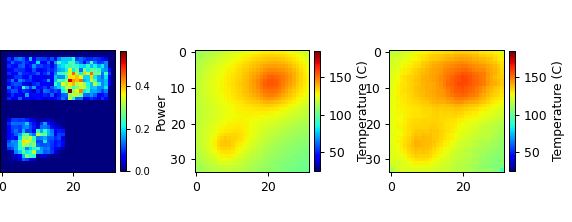}
    \caption{{\em [For an animated version, visit the GitHub repository: https://github.com/asp-dac/asp-dac-1323.git to view the video.]} Video comparing the prediction of transient ThermEDGe against commercial tool-generated temperature contours for T6: (i) left video shows the time-varying power map, (ii) center video shows the commercial tool-generated temperature maps, and (iii) right video shows ThermEDGe-generated temperature maps}
    \label{fig:video}
\end{figure}

Fig.~\ref{fig:video} (left) shows an animated video of the time-varying
power map for T6, where each frame (time-step) is after a 15s
time interval. As before, the corresponding ground truth and predicted
temperature contours are depicted in center and right,
respectively, of the figure.

\noindent
{\bf IREDGe results}
We compare IREDGe-generated contours against the contours generated by ~\cite{pdnsim} across 500 different testcases (10\% of the data, orthogonal to the training set) with varying PDN densities and power distributions. Across the five testcases in Table~\ref{tbl:iredge-qcomm-results}, IREDGe has an average $IR_{err}$ of 0.053mV and a worstcase max $IR_{err}$ of 0.34mV which corresponds to 0.008\% and 0.048\% of VDD respectively. Given that static
IR drop constraints are 1--2.5\% of VDD, a worstcase error of 0.34mV is acceptable in light of the rapid runtimes.  We list the  results of five representative testcases in 
Table~\ref{tbl:iredge-qcomm-results} where the percentage errors in $IR_{err}$ are listed as fraction of VDD$=0.7$V.

\begin{table}[htb]
\caption{Summary of results from IREDGe for 10 different testcases. T16-T20 are testcases which have a chip size that was not in the training set. }
\centering
\label{tbl:iredge-qcomm-results}
\resizebox{\linewidth}{!}{%
\begin{tabular}{||l|l|l||l|l|l||} 
\hhline{|t:===:t:===:t|}
\multicolumn{3}{||c||}{{\bf Chip size: 34x32
}} & \multicolumn{3}{|c||}{{\bf Chip size: 68x32}} \\  
\hhline{|:===::===:|}
\textbf{\#Testcase}  & \textbf{Avg. $\bf IR_{err}$} & \textbf{Max $\bf IR_{err}$} & \textbf{\#Testcase}  & \textbf{Avg. $\bf IR_{err}$} & \textbf{Max $\bf IR_{err}$} \\
\hline \hline
T11 &  0.052mV (0.007\%)  & 0.26mV (0.03\%) & T16 & 0.035mV (0.005\%) & 0.16mV (0.02\%)\\  \hline
T12 &  0.074mV (0.011\%)  & 0.34mV (0.05\%) & T17 & 0.054mV (0.008\%) & 0.42mV (0.06\%)\\ \hline
T13 &  0.036mV (0.005\%)  & 0.21mV (0.03\%) & T18 & 0.035mV (0.005\%) & 0.35mV (0.05\%)\\ \hline
T14 &  0.053mV (0.008\%)  & 0.24mV (0.03\%) & T19 & 0.068mV (0.010\%) & 0.22mV (0.03\%)\\ \hline
T15 &  0.051mV (0.007\%)  & 0.23mV (0.03\%) & T20 & 0.061mV (0.009\%) & 0.38mV (0.05\%)\\
\hhline{|b:===:b:===:b|}
\end{tabular}
}
\end{table}

\begin{figure}[htb]
\centering
\includegraphics[width=9cm]{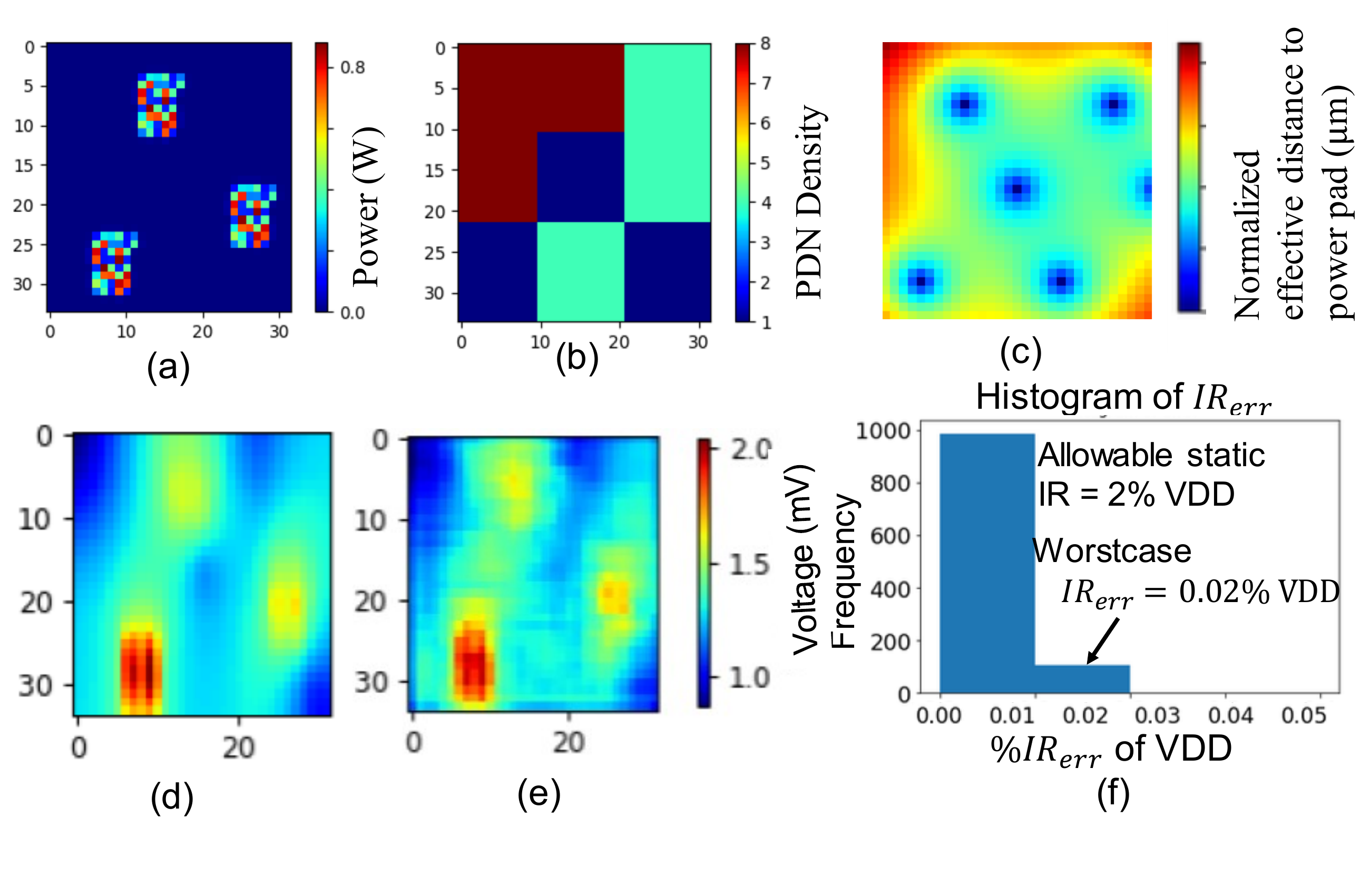}
\vspace{-3em}
\caption{IREDGe static IR drop estimation on T11: (a) input power map, (b) input PDN density map, (c) effective distance to power pad map (d) ground truth IR drop map, (e) predicted IR drop map, and (f)  histogram of $IR_{err}$ showing a worstcase error of 0.16mV.}
\label{fig:ir-result}
\end{figure}

\begin{figure}[htb]
\centering
\vspace{-1.0em}
\includegraphics[width=8.5cm]{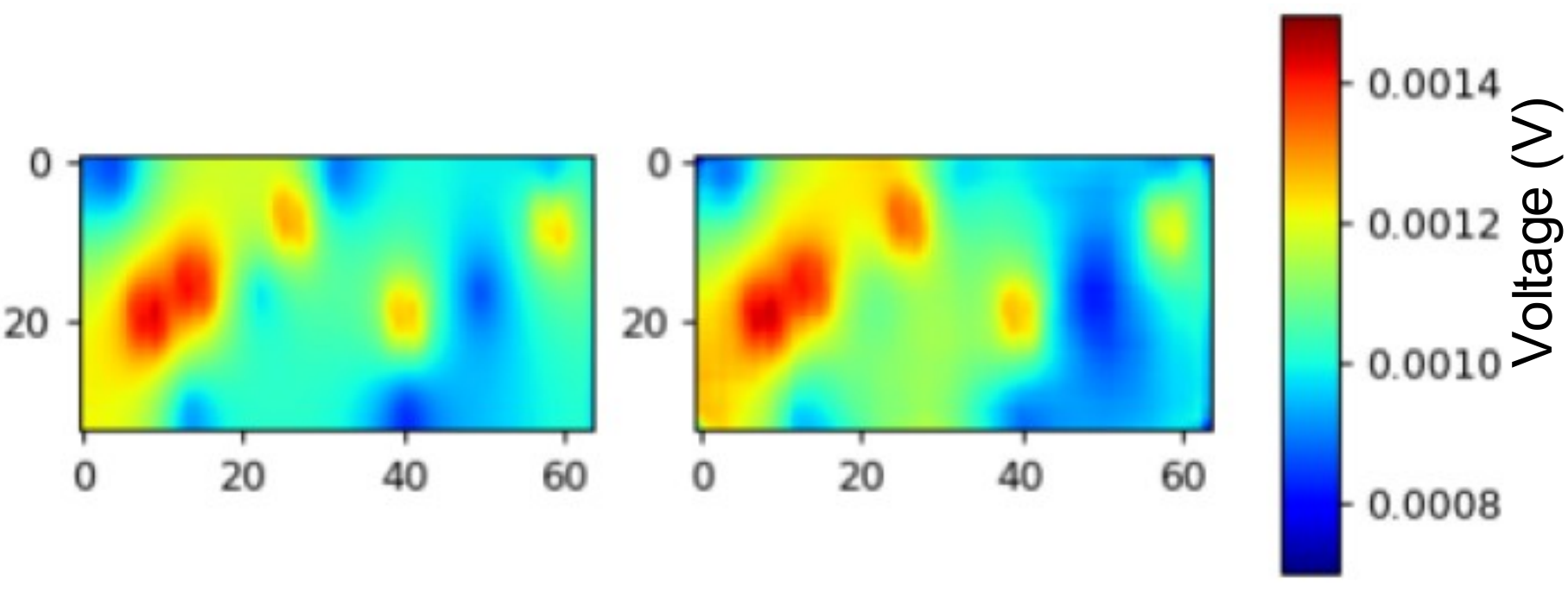}
\caption{Size independent nature of IREDGe: Comparison between (a) Actual IR drop contours and (b) IREDGe-predicted contours for a power map (T16) with size 68$\times$32 using a model that 
was trained on images of size 34$\times$32.}
\label{fig:size-independent}
\vspace{-0.5em}
\end{figure}

A detailed view of T11 is shown in Fig.~\ref{fig:ir-result}. 
It compares the IREDGe-generated IR drop contour plots
against contour plot generated by~\cite{pdnsim}.
The input power maps, PDN density maps, and effective distance to power pad maps are shown
in Fig.~\ref{fig:ir-result}(a), (b), and (c) respectively.
Fig.~\ref{fig:ir-result}(d) and (e) shows the comparison between ground truth and predicted value for the corresponding inputs. It is evident that the plots are similar; numerically, the histogram in Fig.~\ref{fig:ir-result}(f)
shows the \%$IR_{err}$ where the worst \%$IR_{err}$ is less than 0.02\% of VDD.

\noindent
{\bf Size-independence}
One of the primary advantages of using IREDGe for static IR
estimation is that its fully-convolutional nature enables the use of input
images of any size, and the size of the hotspot determines the model rather
than the size of the chip. Since the trained model comprises only of the
trained weights of the kernel, the same kernel can be used to predict the
temperature contours of chip of any size as long as resolution of the represented image remains the same. We test static IREDGe on chips of a different size (T16 -- T20), using a
power distribution of size $68\times32$ as input.
Fig.~\ref{fig:size-independent}(a) compares the actual IR drop of T16 (Fig.~\ref{fig:size-independent}(a)) and the IREDGe-predicted (Fig.~\ref{fig:size-independent}(b)) solution of T16 using a model which was trained on $34\times32$ power maps.
We summarize the results for the rest of the testcases in Table~\ref{tbl:iredge-qcomm-results}.

\noindent
{\bf Runtime analysis} A summary of the runtime comparison of our ML-based EDGe network approach against the temperature and IR drop golden solvers is listed in Table~\ref{tab:runtime}. The runtimes are reported on a NVIDIA GeForce RTX 2080Ti GPU.  With the millisecond inference times, and the transferable nature of our trained models, the one-time cost of training the EDGe networks is easily amortized over multiple uses within a design cycle, and over multiple designs.

\begin{table}[ht]
\centering
\caption{Runtime comparison between EDGe networks and golden thermal analysis and IR drop analysis tools}
\label{tab:runtime}
\resizebox{0.4\textwidth}{!}{%
\begin{tabular}{|l|l|l|l|l|}
\hline
\textbf{Analysis type} & \textbf{\# Nodes} & \textbf{\begin{tabular}[c]{@{}l@{}}Design \\ Area \\ (mm$^2$)\end{tabular}} & \textbf{\begin{tabular}[c]{@{}l@{}}Icepak/\\  PDNSim \\  (minutes)\end{tabular}} & \textbf{\begin{tabular}[c]{@{}l@{}}ThermEDGe/\\      IREDGe\\ (milli seconds)\end{tabular}} \\ \hline
Static thermal & 2.0 million & 64 & 30 mins & 1.1 ms \\ \hline
Transient thermal & 2.0 million & 64 & 210 mins & 10 ms \\ \hline
Static IR drop & 5.2 million & 0.16 & 310 mins & 1.1 ms \\ \hline
\end{tabular}%
}
\end{table}

\begin{figure}[ht]
\centering
\vspace{-0.5em}
\includegraphics[width=8.5cm]{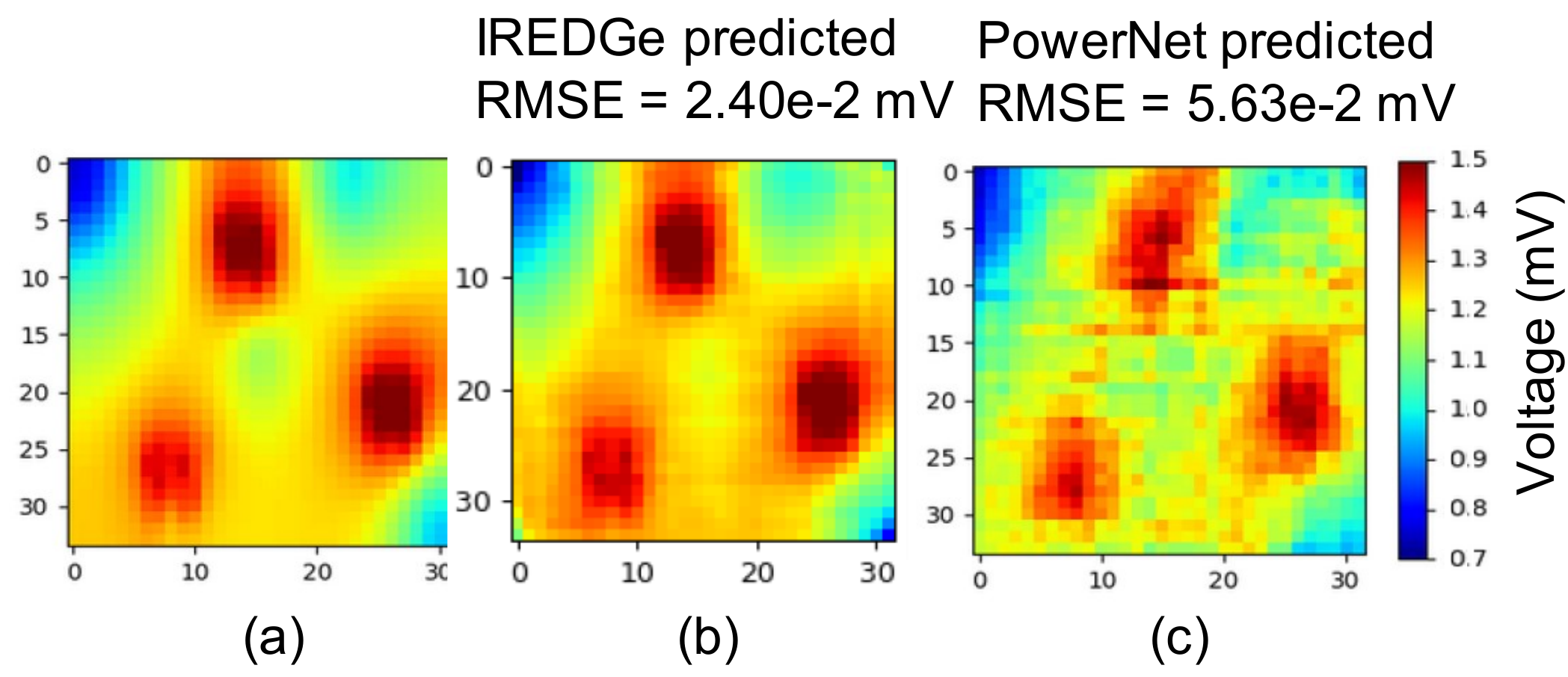}
\caption{IR drop comparisons on T21: (a) ground truth, (b) from IREDGe, and (c) from our implementation of PowerNet.}
\vspace{-1.5em}
\label{fig:powernet-comp}
\end{figure}


\subsection{IREDGe compared with PowerNet}
\noindent
We compare the performance of IREDGe against our implementation of
PowerNet, based on its description in~\cite{powernet}. 
The layout is divided into tiles, and the CNN features are the
2-D power distributions (toggle rate-scaled switching and internal power, total
power, and leakage power) within each tile and in a fixed window of surrounding tiles.
The trained CNN is used to predict the IR drop on a tile-by-tile basis by
sliding a window across all tiles on the chip. The work uses a tile size of
5$\mu$m$\times$5$\mu$m and takes into consideration a 31$\times$31 tiled
neighborhood (window) power information as features. 
For a fair comparison, we train IREDGe under a fixed PDN density and fixed power pad locations that is used to train PowerNet. Qualitatively, IREDGe is superior on three aspects: \\
(1) {\em Tile and window size selection:} It is stated in~\cite{powernet} that 
when the size of the tile is increased from
1$\mu$m$\times$1$\mu$m to 5$\mu$m$\times$5$\mu$m and the size of the resulting window is increased
to represent 31$\times$31 window of 25$\mu$m$^2$ tiles instead of 1$\mu$m$^2$ tiles, the 
accuracy of the PowerNet model improves. 
In general, this is the
expected behavior with an IR analysis problem where the accuracy increases as
more global information is available, until a certain radius after which the
principle of locality  holds~\cite{Chiprout04}. IREDGe bypasses
this tile-size selection problem entirely by providing the entire power map
as input to IREDGe and allowing the network to learn the window size that is
needed for accurate IR estimation.  \\
(2) {\em Run times:} Unlike PowerNet, which trains and infers IR drop on a sliding tile-by-tile
basis, IREDGe has faster training and
inference. IREDGe requires a {\em single} inference, irrespective of the size of the chip while PowerNet performs an inference for every tile in the chip. For this setup and data, it takes 75 minutes to train and implementation of PowerNet, as against 30 minutes for IREDGe. For inference, PowerNet takes 3.2ms while IREDGe takes 1.1ms for a 34$\times$32 chip size. For a chip of $68\times32$ IREDGe takes 1.3ms to generate IR drop contours while PowerNet takes 6.2ms.

(3) {\em Model accuracy:} Since PowerNet uses a CNN to predict IR drop on a
region-by-region basis, where each region is 5$\mu$m by 5$\mu$m,
the resulting IR drop image is pixelated, and the predicted region prediction value does not correlate well with the neighboring regions. 

We compare IREDGe against our implementation of PowerNet on five different testcases T21--25.
These testcases have the same power distribution in T11--15 except that all
the five testscases have identical uniform PDNs, and identical power pad
distributions, as required by PowerNet; IREDGe does not require this.
Fig.~\ref{fig:powernet-comp} shows a comparison between the IR drop solutions
from a golden solver (Fig.~\ref{fig:powernet-comp}(a)), IREDGe
(Fig.~\ref{fig:powernet-comp}(b), and our implementation of
PowerNet (Fig.~\ref{fig:powernet-comp}(c))  for T21 (a representative testcase).
On average, across 
T21--25 IREDGe has an average $IR_{err}$ of 0.028mV and a maximum $IR_{err}$ of 0.14mV as against 0.042mV and 0.17mV respectively for PowerNet.

\section{Conclusion}
\noindent
This paper addresses the compute-intensive tasks of thermal and IR analysis by proposing the use EDGe networks as apt ML-based solutions. Our EDGe-based solution not only improves runtimes but overcomes the window size-selection challenge (amount of neighborhood information required for accurate thermal and IR analysis), that is faced by other ML-based techniques, by allowing ML to learn the window size. 
We successfully evaluate EDGe networks for these applications by developing two ML software solutions (i) ThermEDGe and (ii) IREDGe for rapid on-chip (static and dynamic) thermal and (static) IR analysis respectively. In principle, our methodology is applicable to dynamic IR as well, but is not shown due to the unavailability of public-domain benchmarks.

\bibliographystyle{IEEEtran}
\bibliography{references}

\end{document}